\documentclass[aps,prxquantum,twocolumn,superscriptaddress,longbibliography,nofootinbib,floatfix]{revtex4-2}

\usepackage{graphicx}
\usepackage{amsmath,amssymb,bm}
\usepackage{booktabs}
\usepackage{algpseudocode}
\usepackage[colorlinks=true,allcolors=blue]{hyperref}
\usepackage{url}

\graphicspath{{figures/}}

\newcommand{\ibmfez}{\texttt{ibm\_fez}}
\newcommand{\btheta}{\bm{\theta}}
\newcommand{\bg}{\mathbf g}
\newcommand{\E}{\mathbb{E}}
\newcommand{\RR}{\raggedright\hspace{0pt}}

\begin{document}

\title{Training quantum neural networks one job at a time}

\author{Muhammad Faryad}
\affiliation{Department of Physics, Lahore University of Management Sciences (LUMS),
Lahore 54792, Pakistan}
\affiliation{Computer, Electrical and Mathematical Sciences and Engineering Division (CEMSE),
King Abdullah University of Science and Technology (KAUST), Thuwal 23955-6900, Saudi Arabia}

\author{Rabab Alomairy}
\email[Contact author: ]{rabab.omairy@kaust.edu.sa}
\affiliation{Computer, Electrical and Mathematical Sciences and Engineering Division (CEMSE),
King Abdullah University of Science and Technology (KAUST), Thuwal 23955-6900, Saudi Arabia}

\date{\today}

\begin{abstract}
When a quantum processor is reached through a cloud queue, a variational algorithm pays for
every round trip, and a single round trip can carry many circuits. The classical optimizer
therefore decides how much is learned per job, not only per shot. We study this for a quantum
neural network classifier. We compare optimizers that are given the same number of jobs and the
same shots per job, and that differ only in how they fill each job. This comparison treats hybrid quantum--classical
training as an execution-efficiency problem: how to obtain useful optimization progress from
each costly interaction with a quantum processor. The resulting perspective connects optimizer
design to overhead amortization and efficient use of accelerator resources. A direct port of a batched
pattern search that worked for the quantum approximate optimization algorithm stalls once the
model has a few dozen parameters. What does transfer is the principle behind it: fill each job
with a batch of parameter vectors that is fixed before the job is sent. For circuits built from
Pauli rotations the natural batch is the parameter-shift stencil. We show that its gradient
error falls as the inverse of the shots per job, while the error of simultaneous-perturbation
methods (SPSA) has a floor that no number of shots removes. In simulation, batched
parameter-shift training needs between a fifth and a half of the jobs SPSA needs, for 24 to 48
parameters, and the ordering holds at two noise levels. A new structure-aware stencil search reads the
same kind of rows as sinusoids and also beats SPSA. On IBM's \ibmfez\ processor, in three interleaved runs at six and
eight qubits, the batched optimizer reached SPSA's final training loss within 4 to 20 of its 50
jobs. Its test accuracy was clearly higher at eight qubits; at six qubits the difference was not significant. All hardware numbers can be re-derived from the raw shots.
\end{abstract}

\maketitle

\section{Introduction}
\label{sec:intro}

Variational quantum algorithms run a classical optimizer around a parameterized
circuit~\cite{cerezo2021vqa}. Most studies of these optimizers count circuit evaluations or
shots. On cloud hardware a third quantity matters as much: the number of round trips. Each job
waits in a queue, is loaded onto the control system, and returns its results, and this cost is
paid whatever the job contains~\cite{ravi2021cloud,ravi2021adaptive,menickelly2023latency}. At
the same time a job can hold many circuits. In Qiskit's primitives a single primitive unified
bloc (PUB) can carry an array of parameter vectors for the same
circuit~\cite{javadiabhari2024qiskit}. An optimizer can use that capacity only to the extent that
it can name its next trial points before it sees any result. This is a familiar problem in
high-performance computing (HPC), where work is organized to amortize communication and launch
overheads~\cite{alomairy2022communication,devarakonda2018avoiding}. For a cloud quantum processor,
the question is how to group independent circuit evaluations into each job, so that fewer round
trips are needed to reach a target loss. In HPC terms, the goal is to reach a target solution quality with the
least time and resources. For a variational algorithm, that means grouping independent circuit
evaluations into each submission, and judging them by the progress they buy.

The batched optimizer behind this work was first presented in Ref.~\cite{faryad2026qaoa}, for
the quantum approximate optimization algorithm (QAOA). There, a batched pattern
search~\cite{torczon1997convergence,kolda2003directsearch,hough2001apps} was compared with
COBYLA on \ibmfez\ for portfolio selection with six to twelve assets, with both optimizers given
the same number of jobs and shots. The search sent its whole stencil in one job. After that first
job it already held parameters that COBYLA needed several jobs to match. QAOA at the depths used
there, two to five layers, has only four to ten parameters, so the full stencil fits easily in
one job.

Quantum neural networks (QNNs) are different in two ways. They have tens of parameters even at
small size, so a stencil that probes every coordinate is large. And their loss is an average over
data, so every candidate parameter vector has to be evaluated on a minibatch. A job then holds an
array of (parameter vector) $\times$ (data point) rows, and its shots are shared among all of
them. Whether the batched idea survives these changes is not obvious.

This paper answers that question for a QNN classifier on IBM's \ibmfez\ processor. We hold the
number of jobs and the shots per job fixed, and we ask how each optimizer should fill a job. Our
findings are as follows.

\emph{The literal port fails, the principle transfers.} Compass pattern search moves along one
coordinate per job and stalls at a few dozen parameters. But the principle behind it, a batch
fixed in advance, transfers. For circuits built from Pauli rotations the right batch is the
parameter-shift stencil~\cite{mitarai2018qcl,schuld2019gradients}, which gives exact derivatives
of every model output in one job. The parameter-shift rule and Adam are standard. What is new is
only their job-aware use: the whole stencil goes into a single job.

\emph{A new optimizer, and a picture of all of them.} We introduce a sinusoid stencil search. It
reads the three rows that parameter shift uses along each coordinate as a full sinusoid, moves
every probed coordinate at once, and tests its moves inside the next job, so its line search costs
no extra round trip. Section~\ref{sec:picture} shows, with one figure, how each optimizer fills a
job and what it does with the answer.

\emph{A per-job error analysis explains why.} Under a fixed number of shots per job, the error of
the batched parameter-shift gradient falls as $1/S$, where $S$ is the number of shots in a job.
The error of simultaneous-perturbation stochastic approximation (SPSA)~\cite{spall1992spsa} has a
floor of order $P\Vert\bg\Vert^2$, where $P$ is the number of trainable angles and $\bg$ the
gradient, and this floor does not depend on $S$. For any realistic shot count the batched gradient is far more accurate per
job (Sec.~\ref{sec:theory}).

\emph{Simulation and hardware agree with the analysis.} In simulation, batched parameter-shift
training needs between a fifth and a half of SPSA's jobs. The new sinusoid search, which builds on
the structure that Rotosolve~\cite{ostaszewski2021structure} and NFT~\cite{nakanishi2020smo}
exploit, comes second. On \ibmfez, in three interleaved runs, the batched
optimizer reached SPSA's final training loss within 4 to 20 of its 50 jobs.

The question of what a job buys sits next to a larger literature on shot-frugal
optimization~\cite{sweke2020sgd,kubler2020icans,arrasmith2020rosalin} and on the value of
gradient measurements~\cite{harrow2021gradients}. Those works allocate shots. Here the shots per
job are fixed and the question is which circuits to spend them on, given that a round trip is
the scarce resource. Optimizer comparisons for variational algorithms usually count evaluations
or shots~\cite{lavrijsen2020optimizers,bonetmonroig2023comparison}. We count jobs, which a user
can turn directly into time and cost.

The paper is organized as follows. Section~\ref{sec:jobs} states the job model and the cost of a
job. Section~\ref{sec:model} describes the data, the circuit and its structure.
Section~\ref{sec:optimizers} defines the optimizers, and Sec.~\ref{sec:picture} shows them at work
on a simple landscape. Section~\ref{sec:theory} derives what one job buys each of them. Sections~\ref{sec:sim} and \ref{sec:hw} report simulation and hardware
results. Section~\ref{sec:discussion} discusses limits and what follows. The appendices give the
derivation, the tuning, and the reproducibility checks.

\section{Jobs as the unit of cost}
\label{sec:jobs}

\begin{figure*}[t]
\centering
\includegraphics[width=\textwidth]{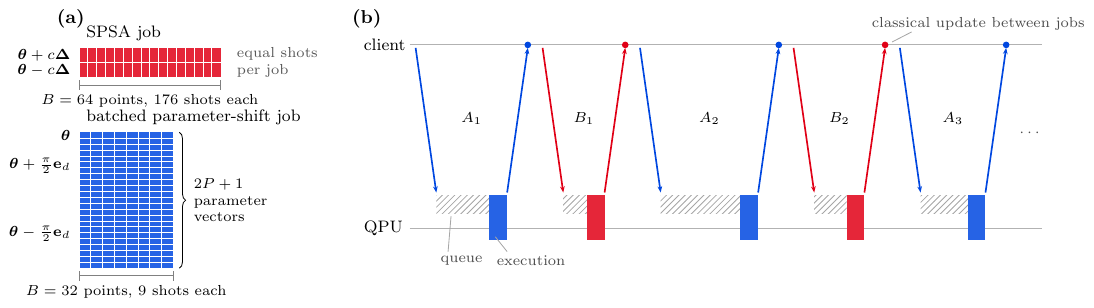}
\caption{The two training arms under a matched job budget. (a)~What one job holds. SPSA sends two
parameter vectors, each on a minibatch of 64 training points. Batched parameter-shift training
sends the current point and its $2P$ shifted copies, each on 32 points. Both jobs carry about
$2.2\times10^4$ shots. (b)~On hardware the two arms are interleaved job by job, so slow drift of
the device cannot favor either one. Every job pays a queue wait and an overhead before it runs. The queue wait varies from job to job
[Fig.~\ref{fig:cost}(b)], and the client sends the next job only after the previous one returns.}
\label{fig:concept}
\end{figure*}

We call a set of circuits submitted together and returned together a \emph{job}. One of our jobs
is one PUB: a circuit with free parameters, and an array of $R$ rows of parameter values, each
row run for $s$ shots. The job's total is $S=Rs$ shots. For an optimizer we write the $k$-th job as
a query set $Q_k$, a list of parameter vectors, each paired with the minibatch it is evaluated on.
We call an optimizer \emph{job-aware} when $Q_k$ depends only on the results of jobs $1,\dots,k-1$.
Every optimizer can be run this way, but they differ greatly in how large a useful $Q_k$ they can
form. COBYLA forms one point. SPSA forms two. A parameter-shift gradient forms $2P+1$.

The cost of a job has three parts. It waits in a queue. It is executed. And it is metered, which
for IBM's job mode means the time the processor is locked to it~\cite{ibmdocs_runtime}. We
measured the metered time of all 306 jobs of this study (three hardware runs of 102 jobs each:
100 training jobs, one read-out job and one drift check) and fit
\begin{equation}
  T_{\rm job} \simeq t_0 + \tau_s S + \tau_r R ,
  \label{eq:cost}
\end{equation}
with $t_0\approx2.0$~s, $\tau_s\approx0.27$~ms per shot and $\tau_r\approx0.5$~ms per row
(Appendix~\ref{app:repro}). Rows are therefore cheap next to shots and fixed overhead. At the
batched arm's size they add 1.2 to 1.6~s to a job. A job of
2{,}336 rows was metered at 9~s and a job of 128 rows at 8~s, for nearly the same number of shots
(Fig.~\ref{fig:cost}). The round trip seen by the user is longer, typically 15 to 30~s on the days
we ran, because of the queue. Job count measures the number of sequential interactions with the
quantum processor, whereas metered time and round-trip time capture distinct execution costs.
Reporting these quantities together helps determine whether improved progress per job translates
into more resource-efficient training.

Two resources thus bound a training run: the number of jobs $J$ and the shots per job $S$. We give
every optimizer the same $J$ and the same $S$. What differs is only how each job is filled.

\section{Model and data}
\label{sec:model}

\paragraph{Data.} We use the hidden-manifold model~\cite{goldt2020hmm} with the generator of the
benchmark suite of Bowles, Ahmed and Schuld~\cite{bowles2024better}. Points live near a
six-dimensional manifold embedded nonlinearly in twelve features, and labels come from a random
network on the manifold coordinates. We draw 300 training and 300 test points once, with a fixed
seed, and standardize features on the training split. An instance with $n$ qubits uses the first
$n$ features, so instances are nested. The task is not hard for classical models: at $n=6$,
logistic regression and an RBF support-vector machine reach about 0.8 test accuracy. That suits our
purpose. We study how fast an optimizer trains a given model, not whether the model beats a
classical one. The classifier is a controlled workload: the model is fixed, and what we measure is
the cost of training it.

\begin{figure*}[t]
\centering
\begin{minipage}[c]{0.60\textwidth}
{\raggedright\small\textbf{(a)}\par}\vspace{2pt}
\centering
\includegraphics[width=\linewidth]{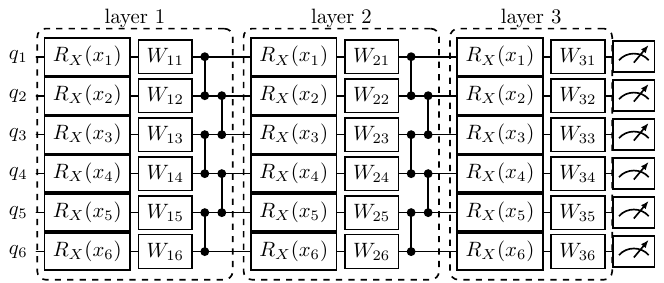}\\[6pt]
\includegraphics[width=0.62\linewidth]{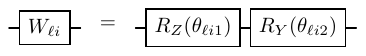}
\end{minipage}\hfill
\begin{minipage}[c]{0.37\textwidth}
{\raggedright\small\textbf{(b)}\par}\vspace{2pt}
\centering
\includegraphics[width=\linewidth]{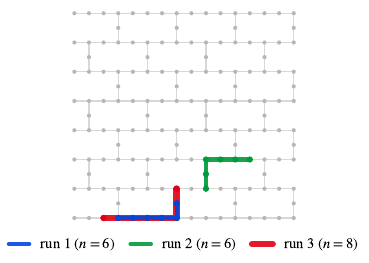}
\end{minipage}
\caption{(a)~The classifier as run at $n=6$ and $L=3$. Features enter through $R_X$ rotations in
every layer. Each block $W_{\ell i}$ holds the two trainable angles of qubit $i$ in layer $\ell$,
so every angle appears once, as a single Pauli rotation. The CZ gates act on neighbors of a
physical chain, first the even pairs and then the odd pairs. The last layer has no CZ gates, since
diagonal gates before a $Z$-basis measurement have no effect. The output is the mean of the six
measured $\langle Z_i\rangle$. (b)~The physical chains used on the 156-qubit heavy-hex \ibmfez\
processor in the three hardware runs. Each chain was chosen at run time as the lowest-error path.}
\label{fig:circuit}
\end{figure*}

\paragraph{Circuit.} The classifier is a data re-uploading
circuit~\cite{perezsalinas2020reuploading,schuld2021fourier} built natively on a chain of
physical qubits (Fig.~\ref{fig:circuit}). Each of $L$ layers applies
\begin{equation}
  U_\ell(\mathbf x,\btheta) = C_\ell\bigotimes_{i=1}^{n} R_Y(\theta_{\ell i 2})\,R_Z(\theta_{\ell i 1})\,R_X(x_i),
  \label{eq:layer}
\end{equation}
where $C_\ell$ applies CZ gates first to the even and then to the odd neighboring pairs of the
chain, and $C_L$ is the identity. The output is the mean single-qubit magnetization,
\begin{equation}
  f(\mathbf x;\btheta) = \frac1n\sum_{i=1}^{n}\langle Z_i\rangle ,
  \label{eq:readout}
\end{equation}
and the loss on a point $(\mathbf x,y)$ with $y=\pm1$ is the logistic loss
$\ell = T^{-1}\log\!\big(1+e^{-Tyf}\big)$ with $T=4$. The model has $P=2nL$ trainable angles and
exactly $(n-1)(L-1)$ CZ gates, with no routing. Our main instance has $n=6$ and $L=3$, so $P=36$ and
10 CZ gates. The second hardware instance has $n=8$, $P=48$ and 14 CZ gates. A noiseless training
scan over $n$, $L$, encoding scale and readout fixed these choices (Appendix~\ref{app:tuning}). At
$n=6$, $L=3$ a fully trained noiseless model reaches about 0.80 test accuracy, level with the classical
references.

\paragraph{Structure.} Every trainable angle enters the circuit once, as a rotation
$e^{-i\theta G/2}$ with $G$ a Pauli operator. Holding the other angles fixed, every expectation
value is then a first-order trigonometric polynomial in that angle. For every data point,
\begin{equation}
  f(\mathbf x;\btheta+t\,\mathbf e_d) = a_d(\mathbf x)\cos t + b_d(\mathbf x)\sin t + c_d(\mathbf x).
  \label{eq:sinusoid}
\end{equation}
Two consequences follow. The parameter-shift rule
$\partial_d f = \tfrac12[f(\btheta+\tfrac{\pi}{2}\mathbf e_d)-f(\btheta-\tfrac{\pi}{2}\mathbf e_d)]$
is exact~\cite{mitarai2018qcl,schuld2019gradients,wierichs2022shift}. And three evaluations
along a coordinate fix the whole one-dimensional landscape of every model output, which
Rotosolve and NFT exploit~\cite{ostaszewski2021structure,nakanishi2020smo}. The loss is a
nonlinear function of $f$, so it is not itself a sinusoid. But it can be computed exactly from the
per-point sinusoids. Both facts shape the optimizers below.

\begin{figure*}[t]
\centering
\includegraphics[width=\textwidth]{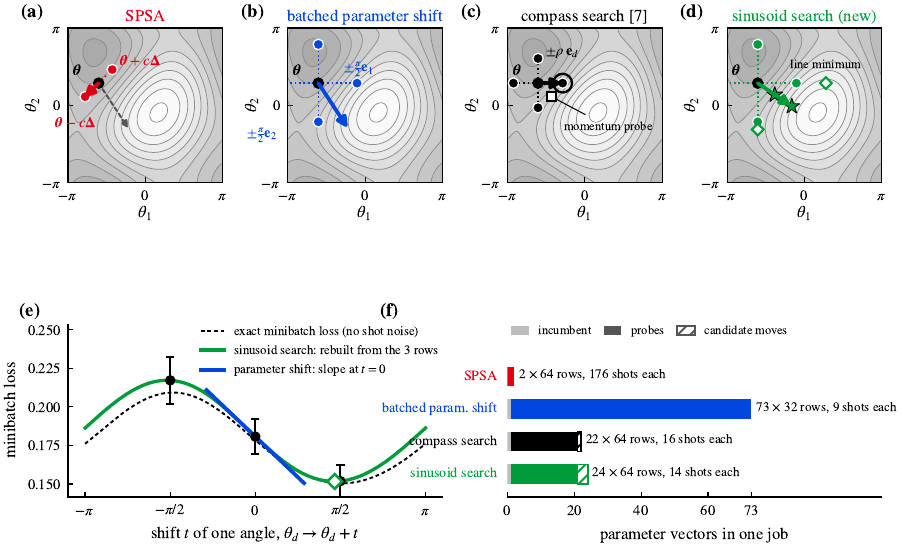}
\caption{How each optimizer uses one job. (a--d)~One job on a toy landscape of two angles.
Lighter shading is lower loss. The black dot is the current point, colored dots are the parameter
vectors sent in the job, and the thick arrow is the move made from the result. Arrow lengths are
for display only. (a)~SPSA probes two points along a random direction $\bm\Delta$ and moves along
$\bm\Delta$. The dashed gray arrow is the true downhill direction. (b)~Batched parameter shift
probes $\pm\pi/2$ along every axis and follows the gradient. (c)~Compass search~\cite{faryad2026qaoa}
probes $\pm\rho$ along each axis plus a momentum probe and moves to the best point (ring) if it
passes a noise test. (d)~The sinusoid search (new) uses the same probes as (b), finds the minimum
along each axis (open diamonds), and sends damped joint moves (stars) as candidates in the next
job. (e)~One angle of the real $n=6$ model at the initial point of run~1, on a minibatch of 64
points. Black dots are the losses measured from the three rows, with 14 shots per row and
standard errors. Parameter shift reads them as a slope at $t=0$ (blue). The sinusoid search
rebuilds the whole loss curve from the same rows (green) and finds its minimum (diamond). The
dashed line is the exact minibatch loss. (f)~The rows of one job of each optimizer at the
settings used in the paper.}
\label{fig:optimizers}
\end{figure*}

\section{Optimizers}
\label{sec:optimizers}

All four optimizers below are written as ask-and-tell loops. ``Ask'' returns the query set of the
next job, and ``tell'' receives the measured outputs $\hat f$ for every row. The rows of a job share
one minibatch, so comparisons inside a job are paired. Table~\ref{tab:optimizers} lists where each
method comes from and marks the one that is new. This interface makes the dependencies explicit.
The rows of one query set can travel together, but the next query set must wait for their
results. The methods differ in how well they use the room in each round trip.

\begin{table*}[t]
\caption{The optimizers compared in this work. Rows per job are given as parameter vectors
$\times$ minibatch points, for $n=6$ ($P=36$) and about $2.2\times10^4$ shots per job.}
\label{tab:optimizers}
\begin{ruledtabular}
\begin{tabular}{p{0.17\textwidth}p{0.22\textwidth}lp{0.21\textwidth}p{0.21\textwidth}}
\RR Optimizer & \RR Origin & Rows per job & \RR What one job tells it & \RR How it moves \tabularnewline
\hline
\RR SPSA & \RR Spall~\cite{spall1992spsa} & $2\times64$ & \RR slope along one random direction & \RR gradient step, decaying gain \tabularnewline
\RR SPSA, 18 directions & \RR Spall~\cite{spall1992spsa} & $36\times16$ & \RR slopes along 18 random directions & \RR averaged step (simulation only) \tabularnewline
\RR Batched parameter shift + Adam & \RR standard rule~\cite{mitarai2018qcl,schuld2019gradients} and Adam~\cite{kingma2014adam}; job-aware use here & $73\times32$ & \RR exact slope along every coordinate & \RR one Adam step \tabularnewline
\RR Compass search & \RR Ref.~\cite{faryad2026qaoa} & $22\times64$ & \RR loss at $\pm\rho$ along 10 coordinates & \RR to the best probe, if it passes a noise test \tabularnewline
\RR Sinusoid stencil search & \RR \textbf{new in this work} & $24\times64$ & \RR full sinusoid along 10 coordinates & \RR all 10 at once, checked in the next job \tabularnewline
\end{tabular}
\end{ruledtabular}
\end{table*}

\paragraph{Batched parameter shift with Adam.} A job holds the current point and its $2P$
shifted copies, each on a minibatch $\mathcal B$ of $B=32$ points. The gradient estimate follows
the chain rule through the loss,
\begin{equation}
  \hat g_d = \frac1B\sum_{i\in\mathcal B}\ell'\big(\hat f_i(\btheta)\big)\,
  \frac{\hat f_i(\btheta+\tfrac{\pi}{2}\mathbf e_d)-\hat f_i(\btheta-\tfrac{\pi}{2}\mathbf e_d)}{2},
  \label{eq:psgrad}
\end{equation}
where $\ell'$ is the derivative of the loss with respect to $f$. It is passed to
Adam~\cite{kingma2014adam} with learning rate 0.1. One job is one Adam step. Both ingredients are
standard. The job-aware part is that all $2P+1$ parameter vectors are known before the job is sent,
so they travel together.

\paragraph{SPSA.} A job holds two points, $\btheta\pm c_k\bm\Delta_k$, with $\bm\Delta_k$ a random
vector of $\pm1$ entries, each on a minibatch of $B=64$ points. The estimate is
\begin{equation}
  \hat\bg = \frac{\hat L(\btheta+c_k\bm\Delta_k)-\hat L(\btheta-c_k\bm\Delta_k)}{2c_k}\,\bm\Delta_k ,
  \label{eq:spsa}
\end{equation}
and the update is $\btheta\leftarrow\btheta-a_k\hat\bg$ with the standard gain sequences
$a_k=a/(k+1+A)^{0.602}$ and $c_k=c/(k+1)^{0.101}$~\cite{spall1992spsa}. We tuned $a$, $c$ and $B$
on a grid in simulation and kept $a=40$, $c=0.2$, $A=10$. A batched variant that averages $K$
perturbation pairs inside one job was also tuned. It appears in simulation only.

\paragraph{Compass pattern search.} This is the batched optimizer of
Ref.~\cite{faryad2026qaoa}. A job holds the incumbent, the points $\btheta\pm\rho\,\mathbf e_d$ for a
block of $k=10$ coordinates, and one momentum probe. The incumbent moves to the best point only if
the improvement beats a multiple of its paired standard error. The radius $\rho$ grows after a
move and shrinks after a failure.

\paragraph{Sinusoid stencil search (new).} This optimizer is introduced here. It keeps the compass stencil but reads it through
Eq.~\eqref{eq:sinusoid}. With probes at $t=0,\pm\pi/2$ on coordinate $d$, the three outputs of every
data point give $a_d$, $b_d$ and $c_d$ in closed form. The minibatch loss along that coordinate,
$\bar\ell_d(t)=B^{-1}\sum_i\ell\big(a_{d,i}\cos t+b_{d,i}\sin t+c_{d,i}\big)$, is then minimized
classically on a fine grid. Each job probes $k=10$ coordinates. It also carries three candidate
moves built from the previous job: the full parallel (Jacobi) step damped by 0.25 and 0.5, and the
best single-coordinate step. These are tested against the incumbent on the shared minibatch, with
the same acceptance rule as above. Rotosolve and NFT update one coordinate per
evaluation~\cite{ostaszewski2021structure,nakanishi2020smo}, and Parrish \emph{et al.} sweep
clusters of parameters~\cite{parrish2019jacobi}. What is specific here is that every probed
coordinate moves at once, the loss is minimized through a nonlinear link, and the line search is
folded into the next job instead of costing a round trip of its own.
Appendix~\ref{app:sinusoid} gives the algorithm.

\section{The optimizers at a glance}
\label{sec:picture}

Figure~\ref{fig:optimizers} shows what each optimizer puts into a job and what it does with the
answer. Panels (a)--(d) use a landscape of two angles, so the geometry can be seen. Panel (e) uses
the real model.

The rule shared by all four is simple. Every row of a job must be chosen before the job is sent.
SPSA chooses two points along one random direction [Fig.~\ref{fig:optimizers}(a)]. Their
difference gives the slope along that direction only, so the step follows $\bm\Delta$ and not the
gradient. With $P$ angles, a random direction makes an angle with the gradient whose cosine is
typically of order $1/\sqrt P$. Most of each SPSA step is therefore spent across the gradient, and
the error of Sec.~\ref{sec:theory} is the price of that.

Batched parameter shift fills the job with the current point and two shifted copies per angle
[Fig.~\ref{fig:optimizers}(b)]. No row depends on the result of another row, so all $2P+1$ can be
sent together, and the full gradient comes back in one round trip. Compass
search~\cite{faryad2026qaoa} also sends a fixed stencil, of $\pm\rho$ probes along a block of
coordinates and one momentum probe [Fig.~\ref{fig:optimizers}(c)]. But it can move to only one of
them per job. In QAOA, with four to ten parameters, that was enough. At $P=36$ it is not.

The sinusoid search sends the same probe geometry as parameter shift but reads it differently
[Fig.~\ref{fig:optimizers}(d,e)]. Because each angle enters once as a Pauli rotation, the output of
every data point is a sinusoid in that angle [Eq.~\eqref{eq:sinusoid}], and three rows fix it.
Panel (e) shows this for one angle of the 6-qubit model. From three noisy losses, parameter shift
keeps only the slope at the current point. The sinusoid search rebuilds the whole curve, which
follows the exact loss to within the shot noise, and finds its minimum without another
measurement. It then moves all probed angles at once, damped, and checks those moves on a fresh
minibatch in the next job. A line search that would normally cost extra round trips is folded into
a job that is sent anyway. Panel (f) shows how full each job is at the settings used below.

Figure~\ref{fig:optimizers} also shows that a larger batch helps only if its rows support a good
update. Compass search sends 22 vectors per job but moves along one coordinate. Parameter shift
sends 73 and moves along all 36. What a job buys depends on how it is filled and on what is read
from it.

\section{What one job buys}
\label{sec:theory}

We now compare the gradient information that a single job gives the two main optimizers. Fix the
current point and a minibatch, and write $\bg$ for the exact minibatch gradient. A row run for $s$
shots returns $\hat f_i=f_i+\varepsilon_i$, where $\varepsilon_i$ has zero mean and variance
$\sigma_i^2/s$. Here $\sigma_i^2$ is the single-shot variance of the readout in
Eq.~\eqref{eq:readout}, at most one. Define
\begin{equation}
  \kappa = \frac1B\sum_{i\in\mathcal B}\ell'(f_i)^2\,\sigma_i^2 ,
\end{equation}
the shot noise that one point contributes to the loss, weighted by how much the loss cares about
that point.

\paragraph{Parameter shift.} With $2P+1$ parameter vectors on $B$ points, $s=S/[(2P+1)B]$. To
leading order in $1/s$, each component of Eq.~\eqref{eq:psgrad} has variance $\kappa/(2Bs)$.
Summing over the $P$ components (Appendix~\ref{app:theory}),
\begin{equation}
  \E\Vert\hat\bg_{\rm PS}-\bg\Vert^2 \simeq \frac{P(2P+1)\,\kappa}{2S}.
  \label{eq:mseps}
\end{equation}
The estimator is unbiased, and its error falls as $1/S$.

\paragraph{SPSA.} With two vectors on $B$ points, $s=S/(2B)$. Even with exact losses the estimator
has an error, because a single random direction is used to stand in for the whole gradient. For
small $c$, $\hat\bg\approx(\bm\Delta\!\cdot\!\bg)\bm\Delta$, and averaging over $\bm\Delta$ gives
$\E\Vert\hat\bg-\bg\Vert^2=(P-1)\Vert\bg\Vert^2$, the familiar dimension factor of random-direction
methods~\cite{spall1992spsa,nesterov2017random,ghadimi2013zeroth}. Shot noise adds a second term,
so that
\begin{equation}
  \E\Vert\hat\bg_{\rm SPSA}-\bg\Vert^2 \simeq (P-1)\Vert\bg\Vert^2 + \frac{P\,\kappa}{c^2 S}.
  \label{eq:msespsa}
\end{equation}
The first term is a floor. No number of shots in the job removes it.

\begin{figure}[t]
\centering
\includegraphics[width=\columnwidth]{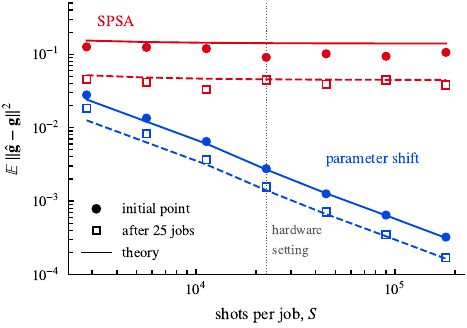}
\caption{Error of one job's gradient estimate against the shots in the job, for the $n=6$ model at
the initial point (filled) and at the point reached after 25 hardware jobs of run 1 (open). Markers
are sampled averages and lines are Eqs.~\eqref{eq:mseps} and \eqref{eq:msespsa}. The minibatch and
parameters are fixed, and the noise model is the mild surrogate of Sec.~\ref{sec:sim}.}
\label{fig:theory}
\end{figure}

\paragraph{Consequences.} The two errors are equal at
$S^\ast\simeq P(2P+1)\kappa/[2(P-1)\Vert\bg\Vert^2]$. In our model $S^\ast$ is a few hundred shots
both at the initial point and partway through training. Above it the batched gradient is the more
accurate one, and its advantage grows with $S$. At the shots per job used on hardware, the relative
error $\E\Vert\hat\bg-\bg\Vert^2/\Vert\bg\Vert^2$ is of order one for parameter shift and of order
$P$ for SPSA. Figure~\ref{fig:theory} checks Eqs.~\eqref{eq:mseps} and \eqref{eq:msespsa} against
direct sampling. Parameter shift follows the $1/S$ law over nearly two decades of $S$. At the
lowest shot counts the sampled error lies above the prediction, by up to about 50\% partway
through training, because Eq.~\eqref{eq:mseps} drops the noise of the incumbent row. SPSA sits
on its floor, somewhat below the linear prediction, because at $c=0.2$ the perturbation
$c\sqrt{P}\approx1.2$~rad is not small and the difference quotient flattens.

How this error turns into progress depends on the optimizer. For a gradient step with a tuned
step size, the expected decrease of a smooth loss scales as
$\Vert\bg\Vert^4/(\Vert\bg\Vert^2+\E\Vert\hat\bg-\bg\Vert^2)$~\cite{bottou2018optimization}. This
predicts a large per-job advantage for parameter shift whenever $S\gg S^\ast$. Two effects make the
realized advantage smaller. Minibatch sampling adds a variance shared by both methods. And both
optimizers average over steps, through Adam's moments or SPSA's decaying gains. The analysis
therefore predicts the direction of the effect, and two signatures that we test below. The batched
optimizer should keep improving as $S$ grows, while SPSA should gain much less from extra shots.

\section{Simulation}
\label{sec:sim}

\paragraph{Setup.} Simulations use an exact statevector model of the circuit. It agrees with
Qiskit to machine precision. Device noise enters through a one-parameter surrogate: a global
depolarizing channel of strength $\eta$, then independent readout flips, then finite shots. We
ran two values. $\eta=0.2$ is close to the value, 0.18, that the depolarizing law fitted to
\ibmfez\ in Ref.~\cite{faryad2026qaoa} gives for a circuit with 10 CZ gates. $\eta=0.03$ matches a full Aer noise model of the same device for
these shallow circuits. The hardware turned out to be close to the milder value
(Sec.~\ref{sec:hw}). Every optimizer was tuned on its own, on a grid, before any hardware use
(Appendix~\ref{app:tuning}). The comparisons below use 100 jobs of $2.25\times10^4$ shots and
20 initial points. The metric is the noiseless training loss of the parameters the optimizer holds,
an oracle we can compute exactly at this size.

\begin{figure*}[t]
\centering
\includegraphics[width=\textwidth]{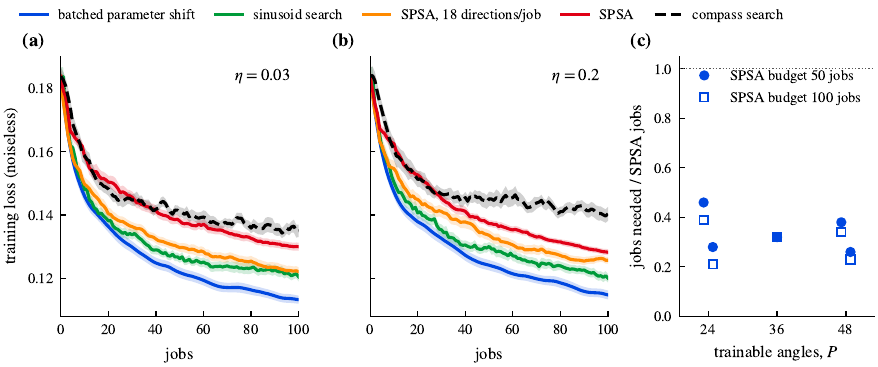}
\caption{Simulated training of the $n=6$, $L=3$ classifier under a matched job budget. (a, b)~Noiseless
training loss of the parameters each optimizer holds, against jobs, for the mild ($\eta=0.03$) and
pessimistic ($\eta=0.2$) noise surrogates. Lines are means over 20 initial points and bands are
standard errors. Every method gets the same shots per job. (c)~Jobs that batched parameter shift
needs to reach SPSA's loss after 50 (filled) and 100 (open) jobs, as a fraction of SPSA's jobs. The
points are for model sizes from $P=24$ to 48, at $\eta=0.2$.}
\label{fig:sim}
\end{figure*}

\paragraph{Comparison.} Figure~\ref{fig:sim}(a,b) shows the result. The ordering is the same at
both noise levels. Batched parameter shift is best, followed by the sinusoid search, SPSA with 18
directions per job, plain SPSA, and compass search. Batched parameter shift reaches SPSA's 50-job
loss in about 20 jobs, and its 100-job loss in 30 to 40. It is ahead of SPSA at job 50 from 19 or
20 of the 20 initial points. The sinusoid search beats SPSA and its batched variant. This is the
pattern-search idea in a form that works for QNNs. It matches batched parameter shift for the
first few jobs and then falls behind.

\paragraph{Why compass search stalls.} Compass search accepts at most one move per job, along one
coordinate. Covering 36 coordinates therefore takes many jobs even when every move succeeds. With
minibatch noise its acceptance test is hard to set. A loose test lets the incumbent follow the
noise. A tolerance of two standard errors or more rejects almost every move near a random initial
point, where the landscape is flat on the scale of the noise. Across all settings we tried,
including the best tolerance of half a standard error, its best loss after 100 jobs stayed near 0.14. Batched parameter shift reaches that
level in about 17 jobs. In QAOA at depths two to five the same search had to cover only four to
ten parameters~\cite{faryad2026qaoa}.

\paragraph{Scaling and shots per job.} The advantage holds from $P=24$ to $P=48$ and for both
$n=6$ and $n=8$ [Fig.~\ref{fig:sim}(c)]. The batched optimizer needs between a fifth and a half of
SPSA's jobs. The largest gap in test accuracy is at $n=8$. In these runs SPSA used $a=20$, a
setting slightly weaker than the $a=40$ used elsewhere (Appendix~\ref{app:tuning}). When the shots
per job are varied from $5.6\times10^3$ to $4.5\times10^4$, SPSA gains from the first doubling about
as much as parameter shift does, but its per-job curve flattens above about $10^4$ shots, while
batched parameter shift keeps improving.
These are the two signatures predicted in Sec.~\ref{sec:theory}. Because SPSA cannot use extra
shots, the batched optimizer also leads per estimated metered second, not only per job.

\section{Hardware}
\label{sec:hw}

\begin{figure*}[t]
\centering
\begin{minipage}{\textwidth}
{\raggedright\small\textbf{(a)}\par}
\centering
\includegraphics[width=0.94\textwidth]{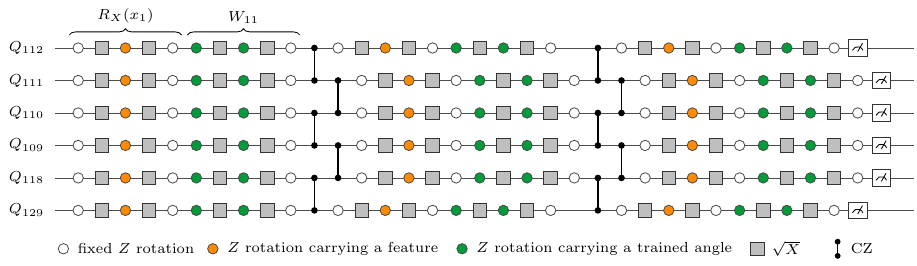}
\end{minipage}\\[10pt]
\begin{minipage}[t]{0.40\textwidth}
{\raggedright\small\textbf{(b)}\par}\vspace{4pt}
\centering
\includegraphics[width=0.92\linewidth]{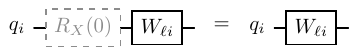}\\[6pt]
\begin{minipage}{0.92\linewidth}\footnotesize
Every feature is set to zero, in every layer. The circuit keeps arm~A's trained
angles and its CZ gates, but its output no longer depends on the data point.
\end{minipage}
\end{minipage}\hfill
\begin{minipage}[t]{0.56\textwidth}
{\raggedright\small\textbf{(c)}\par}\vspace{4pt}
\centering
\includegraphics{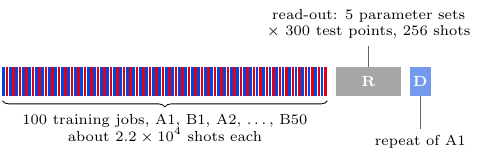}
\end{minipage}
\caption{What was sent to \ibmfez. (a)~The circuit of run~2 after transpilation to the native
gates of the device, on physical qubits $Q_{112}$ to $Q_{129}$ in chain order. Each $R_X(x_i)$ and
each block $W_{\ell i}$ becomes two $\sqrt{X}$ pulses and three $Z$ rotations, so features and
trained angles enter only through $Z$ rotations, which the control system applies as frame
changes. The circuit has 10 CZ gates and a depth of 35. Runs 1 and 3 have the same structure on
other chains. (b)~The encoding-off control. (c)~One hardware run as a sequence of 102 jobs: the
interleaved training jobs of arm A (blue) and arm B (red), one read-out job (R), and a repeat of
the first job of arm A to check for drift (D).}
\label{fig:protocol}
\end{figure*}

\paragraph{Protocol.} We ran batched parameter shift (arm A) against SPSA (arm B) on \ibmfez\ in
three runs: $n=6$ from two initial points, and $n=8$ from one. The runs used job mode on IBM's
Open Plan, without sessions or batches. The arms were interleaved job by job, A1, B1, A2, and so on,
50 jobs each [Fig.~\ref{fig:concept}(b)]. Each job carried about $2.2\times10^4$ shots. Arm A had
up to 7\% fewer, from integer rounding, which works against it. After the 100 training jobs, one
read-out job measured the test set at five parameter sets:
\begin{itemize}
\item the initial point;
\item each arm's final parameters;
\item arm A after only 16 jobs;
\item an encoding-off control, which runs arm A's parameters with every feature set to zero. Its
output carries no information about the data, so its accuracy is the noise floor.
\end{itemize}
A final job repeated arm A's first job unchanged, to bound drift. Dynamical
decoupling~\cite{viola1999decoupling,ezzell2023dd} was on and Pauli twirling~\cite{wallman2016twirling} was off. No other error
mitigation was used. Each run took 102 jobs and 964 metered seconds. The three runs together used
about 48 minutes of processor time. Figure~\ref{fig:protocol} shows the circuit as it was sent,
the control circuit, and the order of jobs in a run.

\begin{figure*}[t]
\centering
\includegraphics[width=\textwidth]{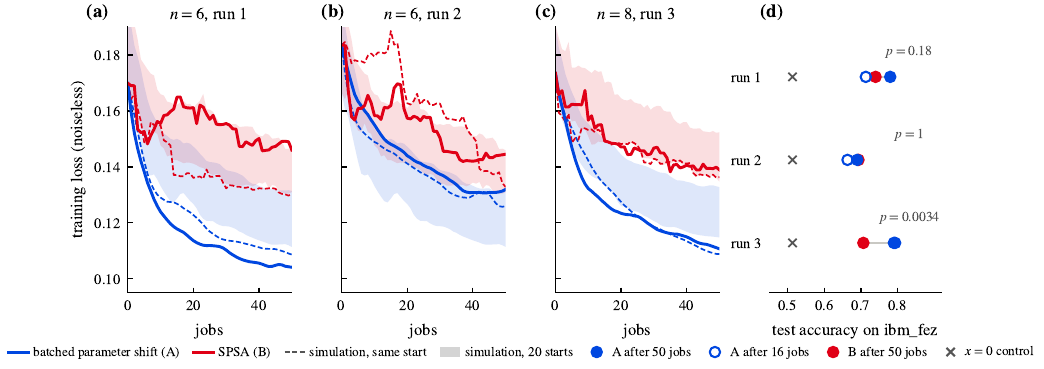}
\caption{Training on \ibmfez. (a--c)~Noiseless training loss of the parameters each arm held after
every hardware job (solid lines). Dashed lines are simulations from the same initial point, and
bands span 20 simulated initial points (10th to 90th percentile, $\eta=0.03$). (d)~Test accuracy
read out on \ibmfez\ for 300 test points. Filled circles are the final parameters of each arm, the
open circle is arm A after 16 jobs, and crosses are the encoding-off control. The $p$ values are
McNemar tests of A against B on the same test points.}
\label{fig:hw}
\end{figure*}

\paragraph{Training.} Figure~\ref{fig:hw}(a--c) shows both arms on the hardware, next to their
simulated twins. In all three runs arm A ends with a lower training loss than arm B. It reaches B's
final loss after 4, 20 and 8 of its jobs. Arm A tracks its simulated twin closely in every run, ending
slightly below it in run 1 and slightly above it in runs 2 and 3. SPSA ends
above its twin in all three runs, clearly in the two $n=6$ runs and only marginally at $n=8$. Its loss climbed for several jobs early in run 1, and a few simulated runs
show similar excursions. Section~\ref{sec:discussion} returns to this.

\paragraph{Test accuracy.} Figure~\ref{fig:hw}(d) shows the read-out. At $n=8$, arm A reached 0.79
and arm B 0.71 ($p=0.003$, McNemar test~\cite{mcnemar1947}), and arm A after only 16 jobs already
beat arm B after 50 ($p=0.01$). The two $n=6$ runs are less clear. Arm A was ahead in run 1 (0.78
against 0.74) and level in run 2. Pooled over all three runs, A answered 130 test points correctly
that B missed, against 93 the other way ($p=0.016$). This pooled test should be read with care. The
runs share the same 300 test points, so the 900 comparisons are not independent, and the result is
carried by run 3. The two $n=6$ runs alone give 80 against 69 ($p=0.41$). The accuracy gain is
therefore clear at $n=8$, unresolved at $n=6$, and smaller and noisier than the gap in training
loss. That is expected on an easy task with 300 test points,
where the two classifiers agree on most points.

\paragraph{Controls.} The encoding-off control sits at the majority-class rate in every run, so the
read-out reflects the data and not an artifact of the readout chain. For the trained parameters, hardware and noiseless
accuracies agree to within 0.02, so device noise did not limit accuracy at these depths. The fitted contraction of the measured output against the ideal one was 0.92 to 0.97.
That is close to the Aer model of the device and far from the pessimistic surrogate. Repeating the
first job at the end changed the outputs by no more than shot noise in runs 1 and 3. In run 2 the
client computer was asleep for about nine hours in the middle of the run, so the last jobs ran
nine hours after the first. There the repeat shows a small systematic shift. Its rms size is
within 1\% of shot noise, but its mean, $-0.009$, is 2.3 standard errors from zero. Because the
arms were interleaved, both were exposed to it equally.

\begin{figure}[t]
\centering
\includegraphics[width=\columnwidth]{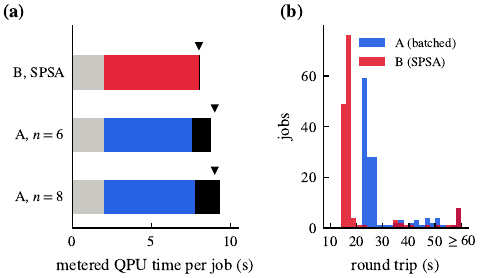}
\caption{What a job costs on \ibmfez. (a)~Metered time per job, split by the fit of
Eq.~\eqref{eq:cost} into fixed overhead (gray), shots (color) and rows (black). Triangles mark the
measured values. (b)~Round-trip time from submission to result for all training jobs of the
three runs. One job delayed by the sleeping client is shown in the last bin.}
\label{fig:cost}
\end{figure}

\paragraph{Cost.} Figure~\ref{fig:cost} breaks down the metered time. A batched job with 2{,}336 or
3{,}104 rows was metered at 9~s, and an SPSA job with 128 rows at 8~s. The round trip is a little
longer for the batched jobs, with medians of 23 to 27~s against 16~s, because their payload is
larger. But the batched arm reached the SPSA arm's final loss within 4 to 20 of its jobs. So it
also wins in wall-clock time and in metered time. Table~\ref{tab:cost} gives the cost of reaching SPSA's final training loss in each run. Arm A
needed 1.8 to 10 times fewer jobs than arm B, 1.6 to 9 times less metered time and 2.2 to 11 times
less round-trip time.

\begin{table}[t]
\caption{Cost of reaching SPSA's final training loss on \ibmfez. Arm A is batched parameter shift
and arm B is SPSA. The target is arm B's noiseless
training loss after its 50th job. Jobs is the first job at which an arm's loss reached the target.
Shots, metered QPU seconds and round-trip seconds are summed over those jobs. The round trip runs
from submission to result and includes the queue. The nine-hour client delay in run~2 came at job
A45, after both arms had reached the target, so it does not enter the table.}
\label{tab:cost}
\begin{ruledtabular}
\begin{tabular}{cccl rrrr}
& & & & & & Metered & Round trip \\
Run & $n$ & Target & Arm & Jobs & Shots & (s) & (s) \\
\hline
1 & 6 & 0.1460 & A & 4 & 84{,}096 & 36 & 118 \\
 &  &  & B & 41 & 923{,}648 & 328 & 1{,}345 \\
\hline
2 & 6 & 0.1445 & A & 20 & 420{,}480 & 180 & 537 \\
 &  &  & B & 36 & 811{,}008 & 288 & 1{,}191 \\
\hline
3 & 8 & 0.1389 & A & 8 & 173{,}824 & 72 & 259 \\
 &  &  & B & 35 & 788{,}480 & 280 & 601 \\
\end{tabular}
\end{ruledtabular}
\end{table}

\section{Discussion}
\label{sec:discussion}

\paragraph{What transfers from QAOA.} The specific algorithm does not transfer, but its principle
does: spend each round trip on a batch that is fixed before the round trip starts. For QAOA at
small depth, a compass stencil is such a batch, and it is small enough to cover every parameter.
For a QNN with tens of parameters and a data average, the compass stencil is the wrong batch. The
parameter-shift stencil is the right one, because each of its rows contributes an exact derivative.
The sinusoid search sits between the two. It keeps the derivative-free form of pattern search and
gains most of the benefit by using the structure of Eq.~\eqref{eq:sinusoid}. In HPC terms, this
is algorithm--system co-design. The optimizer should use what the execution interface makes cheap,
many evaluations per job, and avoid what it makes expensive, sequential round trips. Our results
show this for small QNNs under queued cloud access.

\paragraph{Limits.} The models are small, $n\le8$, so that every claim can be checked against an
exact oracle. The task is easy for classical models, and nothing here concerns quantum advantage.
SPSA was tuned on the surrogate. Its gain depends on the scale of the loss, which device noise
changes. Adam's step does not. SPSA ended clearly above its simulated twins in the two $n=6$
hardware runs. One reading is that it is the more fragile method on hardware, but a short hardware
calibration of its gains would settle whether it was simply mis-tuned. The hardware accuracy
comparison rests on three runs and is significant only at $n=8$. The sinusoid
search was not run on hardware. And all runs used job mode on one device. With reserved access,
through sessions or batches, the queue component of a round trip shrinks, but the metered overhead
in Eq.~\eqref{eq:cost} does not. How large the benefit is depends on the access mode, and on the
relative cost of submission, execution and classical processing. It should be measured again for
other execution environments.

\paragraph{Outlook.} The per-job analysis of Sec.~\ref{sec:theory} applies to any variational
method whose generators are Pauli rotations. It suggests two further steps. Within a job, the shots
could be split unevenly between rows, as shot-frugal methods do between
steps~\cite{kubler2020icans,arrasmith2020rosalin}. And second-order information is available in
the same stencil: the coefficients $a_d$ of Eq.~\eqref{eq:sinusoid} are diagonal curvatures that the
batched job already measures. At larger sizes the job-level comparison can still be simulated, with
GPU statevector or tensor-network simulators, to test how the advantage grows with $P$ beyond the
range that an exact oracle allows. Barren plateaus~\cite{mcclean2018barren,wang2021noisebp} limit
every method here equally. Job-awareness changes how quickly a reachable region is found, not
whether it exists.

\section{Conclusion}

On a queued quantum processor, the classical optimizer decides how much is learned per round trip.
For a quantum neural network built from Pauli rotations, filling each job with the parameter-shift
stencil gives gradients whose error falls with the shots in the job, while SPSA's error does not.
In simulation this cuts the jobs needed by a factor of two to five. On \ibmfez, in three interleaved runs, the
batched optimizer reached SPSA's final loss within 4 to 20 of its 50 jobs, at a metered cost per
job within one second of SPSA's. How the circuits of each job are chosen is therefore a practical
lever on the cost of hybrid training. We think the most useful habit to take from this is simple:
judge optimizers by the quality they reach, the jobs they need and the cost they are metered, not
only by shots.

\begin{acknowledgments}
We thank IBM Quantum for access to \ibmfez\ through the IBM Quantum Open Plan, including a
promotional allocation of 180 minutes of processor time. This work was supported by King Abdullah
University of Science and Technology (KAUST) through baseline research funding. M.F. is a member of
the Qiskit Advocate program. The views expressed are those of the authors and do not reflect the
official policy or position of IBM or the IBM Quantum team. The idea of this work, the extension of
the batched approach from QAOA to QNNs, and all conclusions are the authors' own. Claude (Anthropic)
was used to help write the code and draft the text. The authors verified all of it and take full
responsibility.
\end{acknowledgments}

\section*{Data availability}
All code, the raw shot records of every hardware job, the IBM job identifiers and the cached
simulation results are available at \url{https://github.com/muf148/qnn-job-budget}. A verification
notebook replays both optimizers from the raw shots and reproduces every saved parameter vector and
every reported hardware number. Version 1.0.0 of the repository, which matches this paper, is archived on Zenodo at
\url{https://doi.org/10.5281/zenodo.23970402}. Code is released under the MIT license and data under CC-BY-4.0.

\appendix

\section{Derivation of the per-job errors}
\label{app:theory}

Fix $\btheta$ and a minibatch $\mathcal B$ of $B$ points. Write $f_i^{\pm d}$ for the output of point
$i$ at $\btheta\pm\tfrac{\pi}{2}\mathbf e_d$, and $w_i=\ell'(f_i)$. Measured outputs are
$\hat f=f+\varepsilon$, independent across rows, with $\E\varepsilon=0$ and
$\E\varepsilon^2=\sigma^2/s$.

\paragraph{Parameter shift.} Expanding Eq.~\eqref{eq:psgrad} to first order in the noise,
\begin{equation*}
\hat g_d-g_d = \frac1B\sum_i\Big[w_i\frac{\varepsilon_i^{+d}-\varepsilon_i^{-d}}{2}
 + \ell''(f_i)\,\varepsilon_i\,\partial_d f_i\Big].
\end{equation*}
The second term comes from the noise in the incumbent row. It is shared by all components and is
smaller by a factor $|\ell''\partial_df|/|\ell'|$, which is small for the logistic loss at the
margins reached here. We drop it. The first term has variance
$B^{-2}\sum_i w_i^2\sigma_i^2/(2s)=\kappa/(2Bs)$. Summing over $d$ and using $s=S/[(2P+1)B]$ gives
Eq.~\eqref{eq:mseps}. The sampled errors in Fig.~\ref{fig:theory} include the dropped term and
noise in $\ell'$. The plotted prediction uses the integer number of shots per row that was
actually sent, which matters at the lowest $S$. With it, the sampled error at the initial point is
16\% above the prediction at the lowest shot count. Partway through training it is 30 to 50\%
above at the two lowest shot counts and about 10\% above by $S\approx2\times10^4$.

\paragraph{SPSA.} Write $\delta=\hat L^+-\hat L^-$. Then
$\delta = 2c\,\bm\Delta\!\cdot\!\bg + O(c^3) + \xi$, where $\xi$ is the shot noise of the difference
of two minibatch losses, with variance $2\kappa/(Bs)$ to first order. The estimate is
$\hat\bg=\delta\bm\Delta/(2c)$. For Rademacher $\bm\Delta$, $\E[\bm\Delta\bm\Delta^{\mathsf T}]=I$ and
$\Vert\bm\Delta\Vert^2=P$, so
$\E\Vert(\bm\Delta\!\cdot\!\bg)\bm\Delta-\bg\Vert^2=(P-1)\Vert\bg\Vert^2$. The noise term contributes
$P\,\E\xi^2/(4c^2)=P\kappa/(2c^2Bs)$. With $s=S/(2B)$ this gives Eq.~\eqref{eq:msespsa}. The $O(c^3)$
term is a bias. At $c=0.2$ and $P=36$ the perturbation is not small. The true difference quotient
saturates, which shrinks the spread of the estimate. This is why the sampled SPSA errors in
Fig.~\ref{fig:theory} lie somewhat below the linear prediction.

\paragraph{Minibatch noise.} Both methods also carry the variance of the minibatch gradient around
the full-data gradient. In our model it is comparable to, or larger than, the squared norm of the full-data gradient at $B=32$. It is shared
by both methods, and it is one reason why the realized job savings are smaller than the ratio of
Eqs.~\eqref{eq:mseps} and \eqref{eq:msespsa} suggests.

\section{Tuning and model selection}
\label{app:tuning}

All choices were made in simulation before any hardware run. The model scan trained the noiseless
classifier to convergence with exact gradients for $n\in\{4,6,8\}$, $L\in\{2,3,4\}$, encoding scales
0.5 and 1, and two readouts: mean magnetization and a single central qubit. The mean magnetization
gave the lower loss in every case at $n=6$, and the same or better test accuracy in all but one
case, where it trailed by 0.007. $L=3$ was the shallowest depth that
reached the classical references at $n=6$.

Each optimizer was then tuned at $n=6$, $L=3$, over 100 jobs and six initial points, under the
pessimistic surrogate:
\begin{itemize}
\item batched parameter shift: learning rate $\{0.05,0.1,0.2\}$ and $B\in\{8,16,32,64\}$;
\item SPSA: $a\in\{10,20,40,80\}$, $c\in\{0.05,0.1,0.2\}$, $B\in\{16,64\}$, and averaging over 18
directions per job;
\item compass search: block size, acceptance tolerance, minibatch, and pipelined line probes;
\item sinusoid search: block size, minibatch, step damping, and a trust region.
\end{itemize}
The settings used are listed in Sec.~\ref{sec:optimizers}. For SPSA with one direction per job, $a=40$ with $c=0.2$ was best, closely followed by $a=20$ with $c=0.1$. The hardware arm
and the main comparison used $a=40$, $c=0.2$. The scaling and shots-per-job studies used $a=20$,
$c=0.2$, which ranked fourth. At $\eta=0.2$ it reaches 0.132 after 100 jobs against 0.128 for
$a=40$, so those two panels favor parameter shift slightly.

\section{Sinusoid stencil search}
\label{app:sinusoid}

\begin{figure}[t]
\vspace{2pt}\hrule\vspace{3pt}
\noindent\textbf{Algorithm 1.}~Sinusoid stencil search. One iteration is one job.
\vspace{2pt}\hrule\vspace{4pt}
\begin{algorithmic}[1]
\Require incumbent $\btheta$, block size $k$, damping set $\mathcal A$, tolerance $\tau$
\State $\mathcal V\gets\emptyset$ \Comment{candidates to verify}
\For{each job}
  \State draw a block $D$ of $k$ coordinates and a minibatch $\mathcal B$
  \State submit $\{\btheta\}\cup\{\btheta\pm\tfrac{\pi}{2}\mathbf e_d\}_{d\in D}\cup\mathcal V$ on $\mathcal B$
  \State $\btheta_0\gets\btheta$
  \If{$\mathcal V\neq\emptyset$} \State $\mathbf v^\ast\gets\arg\min_{\mathbf v\in\mathcal V}\hat L(\mathbf v)$
     \If{$\hat L(\mathbf v^\ast)<\hat L(\btheta)-\tau\,\mathrm{SE}$} $\btheta\gets\mathbf v^\ast$ \EndIf
  \EndIf
  \For{$d\in D$} \State fit $a_{d,i},b_{d,i},c_{d,i}$ for every $i\in\mathcal B$ from the three probes
     \State $t_d^\ast\gets\arg\min_t\,\bar\ell_d(t)$ \Comment{classical, on a grid}
  \EndFor
  \State $\mathcal V\gets\{\btheta_0+\alpha\textstyle\sum_{d\in D}t_d^\ast\mathbf e_d\}_{\alpha\in\mathcal A}
          \cup\{\btheta_0+t_{d^\ast}^\ast\mathbf e_{d^\ast}\}$
\EndFor
\end{algorithmic}
\vspace{2pt}\hrule
\end{figure}

Algorithm~1 states the method. From the probes at $0$ and $\pm\pi/2$,
$b_{d,i}=\tfrac12(f_i^{+d}-f_i^{-d})$, $a_{d,i}=f_i-\tfrac12(f_i^{+d}+f_i^{-d})$ and
$c_{d,i}=f_i-a_{d,i}$. The candidates built from one job are verified in the next job, on a fresh
minibatch, so a step costs no extra round trip. SE is the paired standard error of the difference
between candidate and incumbent over the minibatch. We used $k=10$, $B=64$,
$\mathcal A=\{0.25,0.5\}$ and $\tau=0.5$. A job then holds 24 parameter vectors. The best
single-coordinate step $d^\ast$ is the one with the largest predicted decrease.

\section{Reproducibility and verification}
\label{app:repro}

Each hardware run is recorded in two files. The first holds the settings, the parameters held by
each arm after every job, the read-out, the controls, and a ledger with the IBM job identifier,
rows, shots, metered seconds and round-trip time of every job. The second holds every measured
bitstring of every row of every job. From the second run on, the files also record a calibration
snapshot of the qubits used, the software versions, the submitted circuit (QPY and OpenQASM~3),
and the parameter rows and minibatch indices of every job.

The verification notebook replays both optimizers from the raw bitstrings alone, using the same
code and random seeds as the run. It reproduces every saved parameter vector of all three runs to
within $10^{-14}$, including the stored job inputs where they exist. It then recomputes the
read-out accuracies, the McNemar tests and the drift control from the raw read-out shots.

The cost fit of Eq.~\eqref{eq:cost} uses all 306 jobs of the three runs. IBM reports metered time in whole seconds.
Within that resolution, the residuals have a standard deviation of 0.2~s. The 306 jobs come in only
four combinations of shots and rows, and the shot coefficient is fixed mostly by the three read-out
jobs, so the fit summarizes these settings rather than a general cost model. Metered usage was read
from the job record only after the record was final. Earlier in the session it reads zero.

\bibliography{refs}

\end{document}